\documentstyle[twoside,fleqn,espcrc2]{article}

\input epsf

\def\cfig#1{}

\def\myfig#1#2#3{  
\begin{figure}[htb]
\mbox{
  \epsfxsize=70mm \epsfbox{#1}  
}%}
\vspace{-1.1cm}
\caption{#2}
\label{#3}
\vspace{-0.8cm}
\end{figure}
}
 
\newcommand{\figDiagram}{
  \myfig{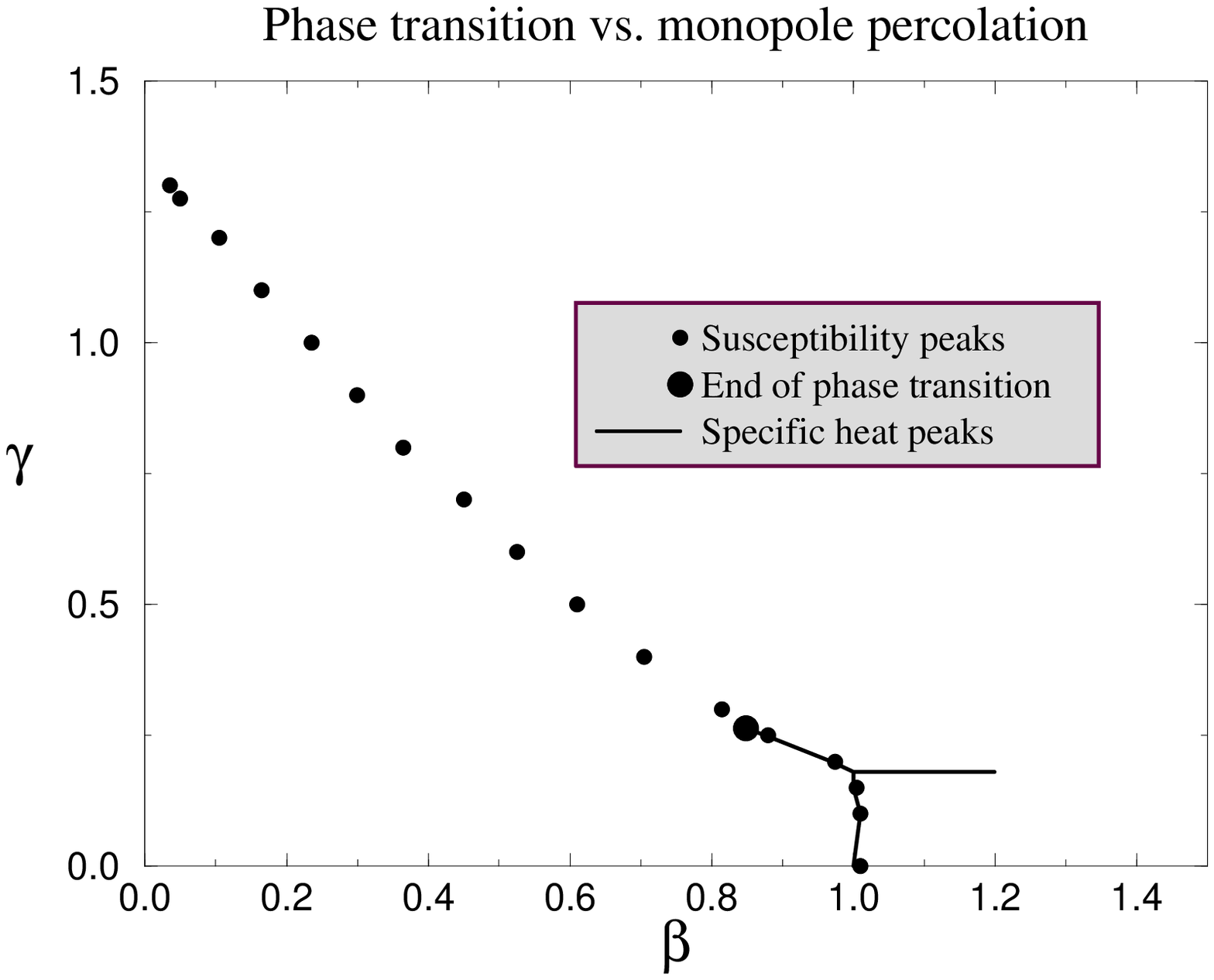}{The phase diagram and the monopole percolation
in SQED}{fig:Diagram}
}

\newcommand{\figHistoa}{
  \myfig{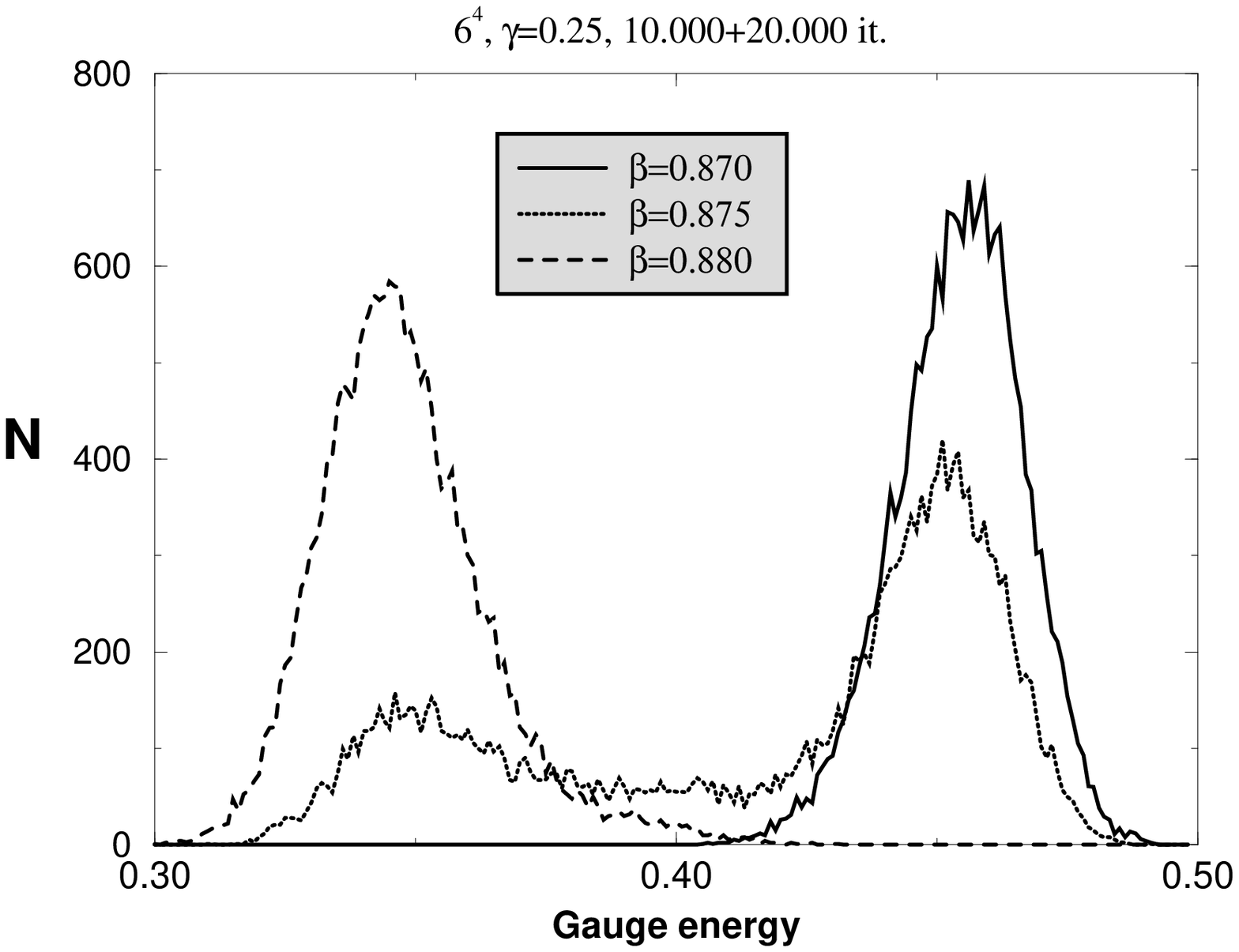}{Histograms of the gauge energy near to the $C_v$ peak}
{fig:Histoa}
}

\newcommand{\figHistob}{
  \myfig{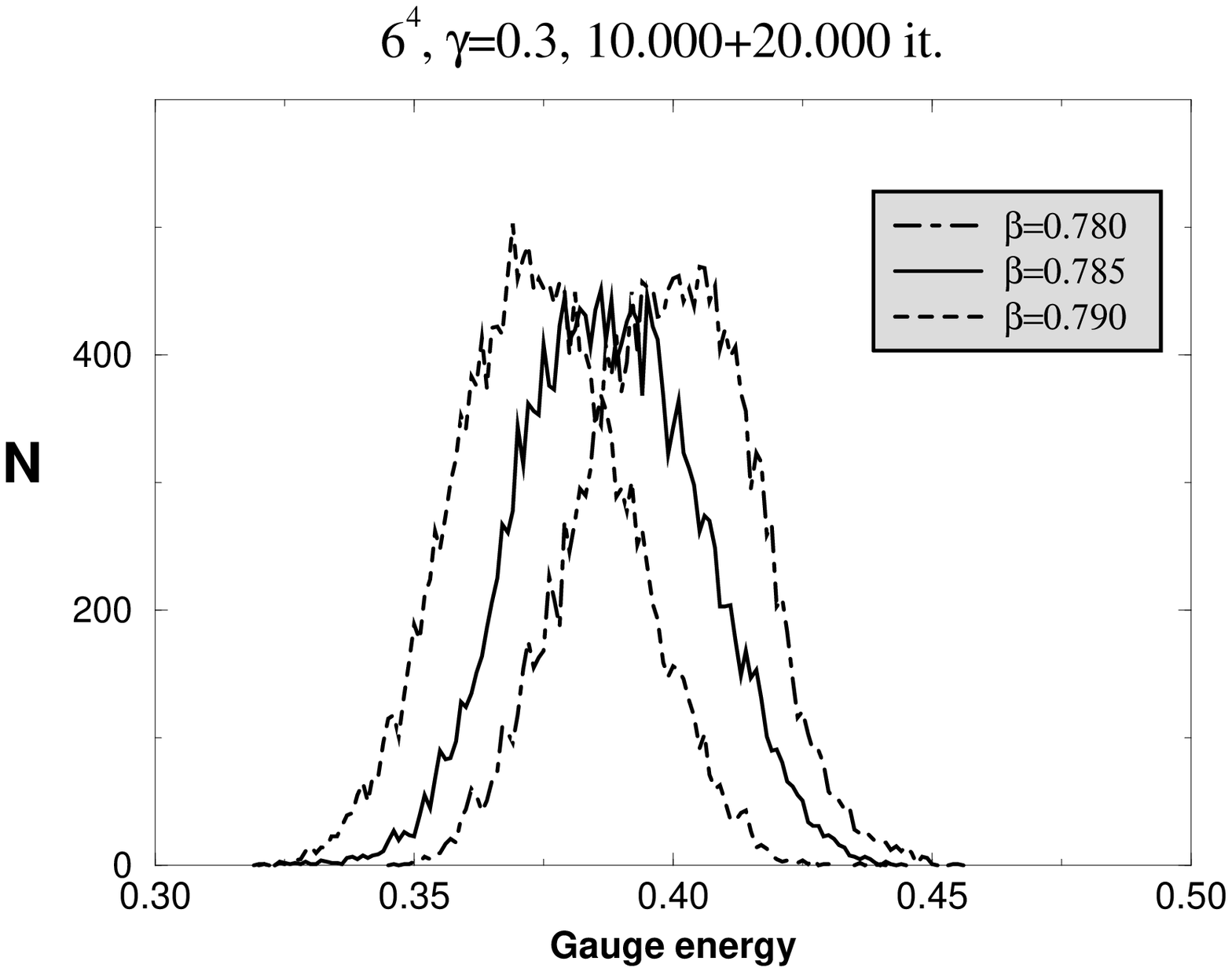}{Histograms of the gauge energy near to the $C_v$ peak}
{fig:Histob}
}

\newcommand{\fignms}{
  \myfig{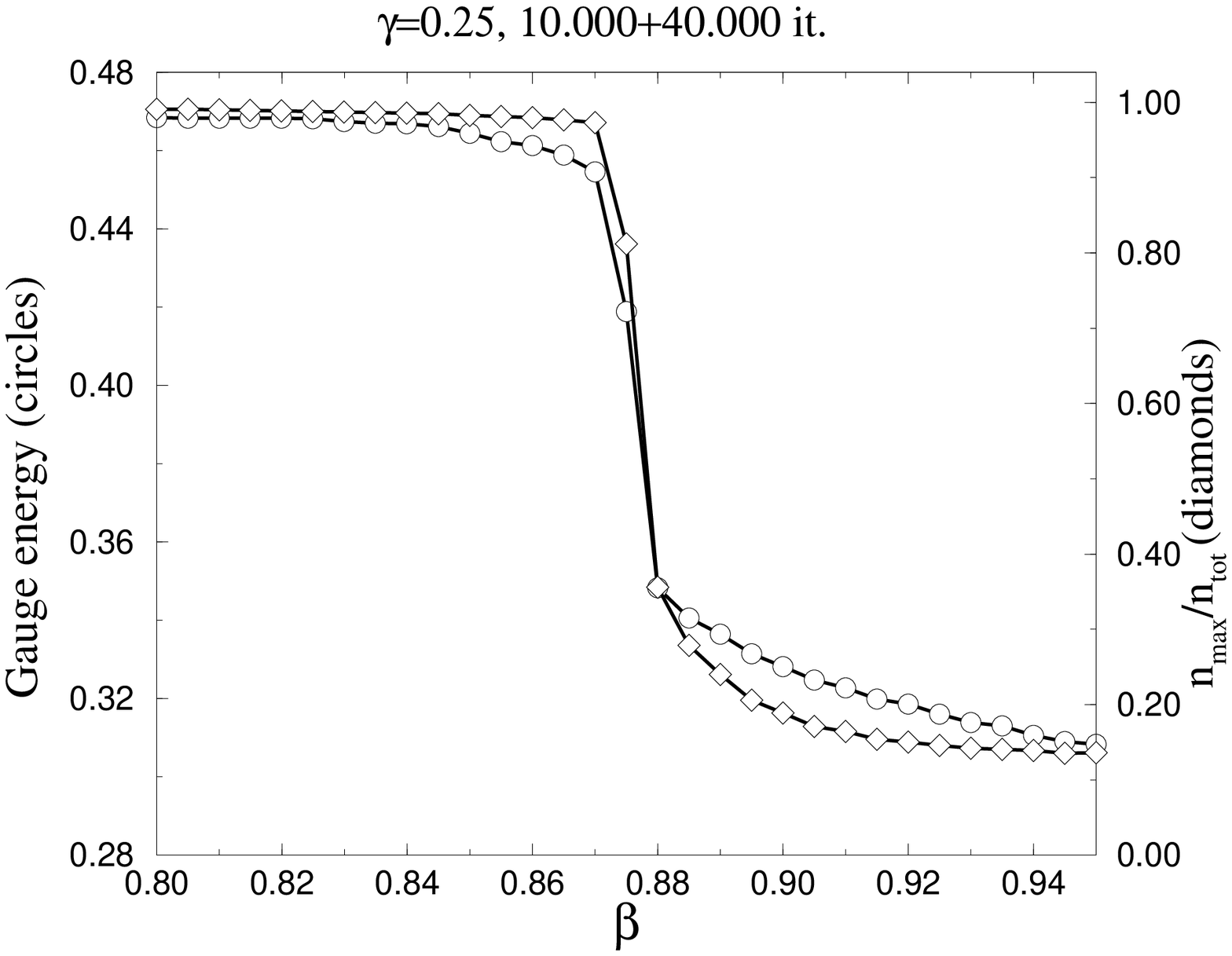}{$n_{max}/n_{tot}$ {\em vs.} gauge energy}
{fig:nms}
}

\newcommand{\figPeak}{
  \myfig{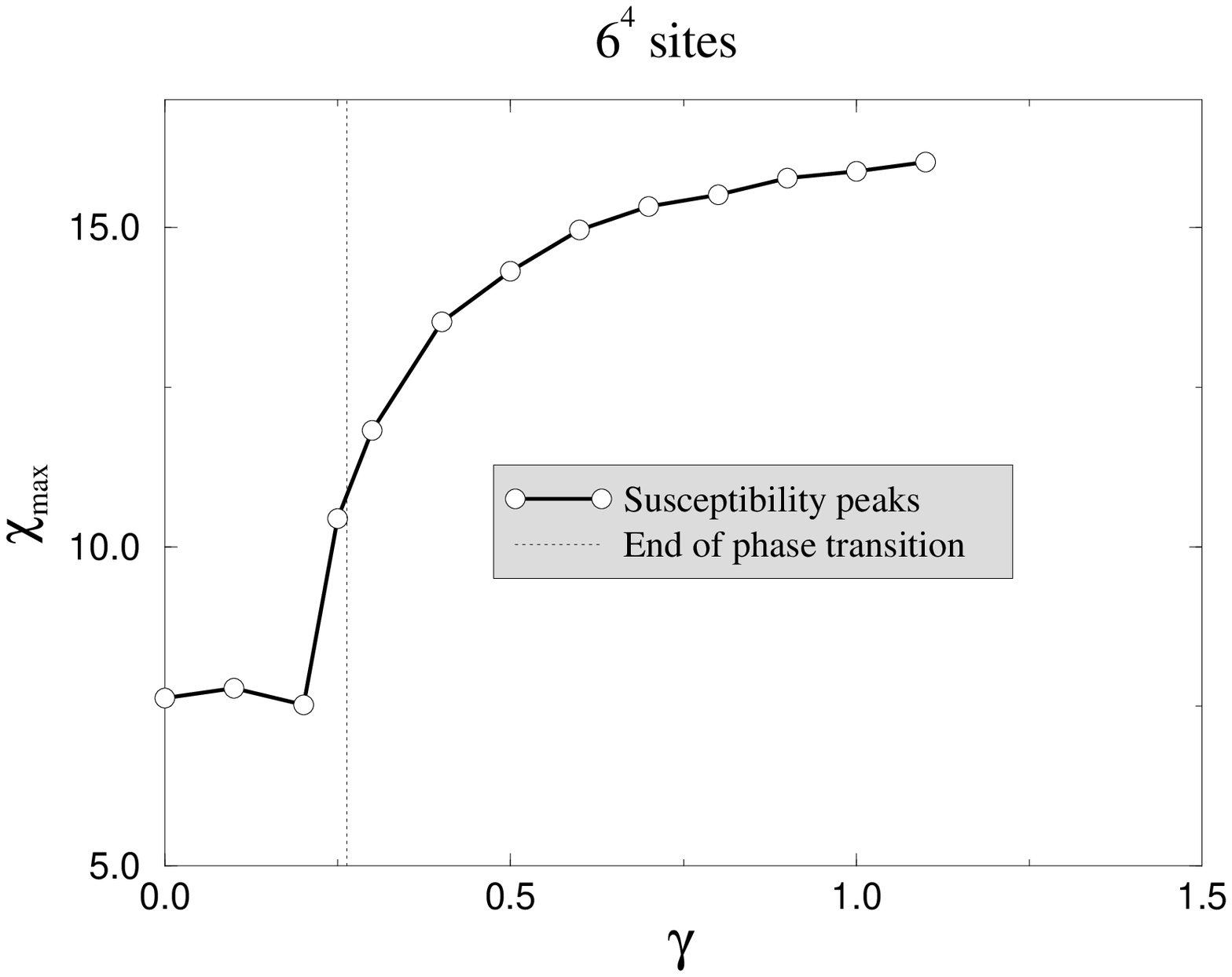}{Susceptibilty peak maximum along the percolation line}
{fig:Peak}
}

\title{ \hfill {\normalsize  UAB--FT/426} \\
 \hfill {\normalsize hep--lat/9709023}\\
Monopole percolation in scalar QED}

\author{M. Baig
        \thanks{e-mail:{\tt baig@ifae.es},
                URL:{\tt http://www.ifae.es/$\sim$baig}}
        and 
        J. Clua 
        \thanks{e-mail:{\tt clua@ifae.es},
                URL:{\tt http://www.ifae.es/$\sim$clua} }
        \\[15pt]
        Institut de F{\'\i}sica d'Altes Energies (IFAE). 
        Facultat de Ci{\`e}ncies. Edifici Cn\\
        Universitat Aut{\`o}noma de Barcelona.
        08193 Bellaterra (Barcelona) Spain
}

\begin{document}

\begin{abstract}
  Monopole Percolation was first introduced in the study of the non-compact
  lattice QED in both, the pure case and coupled to Higgs fields. Monopole
  percolation has been also observed coupled to the monopole condensation in
  the study of the pure gauge compact QED. We present here the results coming
  from the analysis of the role of the monopole percolation in the coupled
  gauge-higgs compact QED.
\end{abstract}


\maketitle

\section{Introduction}
One of the most challenging unsolved questions that arise in the
application of lattice theory to particle physics is to understand
the nature of the phase structure of lattice Quantum Electrodynamics.

Numerical simulations of pure gauge Lattice QED with periodic boundary 
conditions and using the compact gauge action $S=\beta\cos\theta$ show
the existence of a first order phase transition~\cite{DK88}.
This avoids a continuum
limit for the theory. The introduction of matter fields seems not to 
change this behavior.

Dagotto Kogut and Koci{\'c} ~\cite{DKK} 
formulated a version of QED using a non compact form of 
the gauge term $S=-\beta\theta^2$, i.e. keeping only the first term in the 
Taylor development. In this case the full theory exhibits a second order 
phase transition. Nevertheless, the pure gauge non-compact theory, being 
the action Gaussian, is simply trivial.

On the other hand, it has been recognized that the monopoles play a 
central role in the explanation of the phase structure of lattice QED. They
produce {\em disorder} and give rise of the confinement via the dual 
superconductor mechanism. In this sense, the gauge vacuum behaves as 
a {\em magnetic superconductor}, (a monopole condensate). The knowledge
of the behavior of monopoles in an actual simulation may help in the 
comprehension of the phase structure of QED.

\section{Monopole percolation}
In $d=3$ monopoles are point-like excitations, while in $d=4$ they are 
one-dimensional excitations. For this reason, its behavior in a finite 
lattice is far from being trivial. 

An important observation was made by Kogut, Koci\'c and Hands~\cite{KKH92}.
They showed
that in pure gauge non-compact QED monopoles percolate and satisfy the
hyperscaling relations characteristic of and authentic second-order phase 
transition. 

Baig, Fort and Kogut~\cite{BFK94}
showed that in the compact pure gauge theory, just 
over the phase transition point, monopoles {\em condensate} and  
also {\em percolate}. They pointed out that the strange behavior of this 
phase transition -its unexpected first order- 
can be related to the confluency of this two phenomena.

Furthermore, Baig, Fort, Kogut and Kim~\cite{BFKK95}
showed that in the case of non-compact
QED coupled to scalar Higgs fields, the monopole percolation phenomena 
-previously observed over the gauge line- actually propagate into the full
$(\beta-\gamma)$ plane, (being $\gamma$ the gauge-Higgs coupling). 
Surprising, this monopole-percolation phenomenon is decoupled from the 
phase transition line that separates the confined and the Higgs phases,
a transition that is of second order, and logarithmically trivial.

\section{Our analysis}
We have performed a numerical simulation of the compact Lattice QED  
coupled to Higgs fields of unitary norm. We have reproduced some results 
about the phase diagram previously obtained by Alonso et al.~\cite{A93}, but 
measuring at the same time the behavior of monopoles -condensation and 
percolation. Results of this analysis are summarized in Fig.~\ref{fig:Diagram}.

\figDiagram

The comparison of this two figures is very instructive. In the non-compact 
case, the monopole-percolation line remains decoupled from the phase 
transition besides in the compact case, the previously observed confluenze 
of condensation-percolation-transition in the pure gauge case,
still occurs in all the plane $(\beta-\gamma)$. 

An interesting result is that percolation occurs even when the transition 
line finishes. In this case, the behavior of all the 
percolation-related parameters change suddenly. Preliminary results from a 
finite-size scaling of the monopole susceptibility seems to suggest that
the behavior is that of the pure bond percolation (as in the trivial 
non-compact case!)

\section{Analysis of the results}

\begin{itemize}

\item {\underline{Energy histograms}}. In Fig.~\ref{fig:Histoa} we present the histograms
\figHistoa
of the energy  for several values of the gauge coupling, near the $C_v$peak,
 keeping always the Higgs 
coupling fixed $\gamma=0.25$. Lattice size is $6^4$ and the number of 
lattice sweeps is 30.000 per point on a thermal cycle.
The clear two-peaks structure of the histogram 
suggest that this phase transition line is of second order, like the pure 
gauge case.
Fig.~\ref{fig:Histob} \figHistob
collects the same measurements but now keeping $\gamma$=0.3, i.e. just 
above the end-point of the phase transition. The {\em shadow} of the phase 
transition is visible as a little flattening of the histogram, but the 
two-peaks structure has clearly disappeared.

\fignms

\item {\underline{Monopole percolation}}. In Fig.~\ref{fig:nms}
we collect together the 
internal energy and the monopole-percolation parameter $n_{max}/n_{tot}$ 
(number of sites in the larger cluster over the total number of connected 
sites) measured simultaneously over the line $\gamma=0.25$ hat crosses the 
phase transition. Note that the discontinuity of {\bf both} parameters 
occurs at the same point. Since monopole density also decreases 
suddenly at the same point, we can conclude, as in the pure gauge case,
 the concurrence of the three phenomena: Phase Transition, Monopole 
Condensation and Monopole Percolation.

\item {\underline{Monopole susceptibility}}. We have measured the monopole
susceptibility for different values of the Higgs coupling in order to
determine the {\em monopole percolation line} in the phase diagram.
This line remains coupled to the phase-transition line up to its end-point
( see Fig.~\ref{fig:Diagram} ).
Above this point, the percolation-line continues approaching 
the vertical axis. An interesting observation (Fig.~\ref{fig:Peak}) 
is that the value of the 
maximum of the susceptibility changes suddenly over the end-point, i.e.
precisely when the percolation decouples of condensation.
Above the value of $\gamma=1$ the behavior of this peak with the lattice size
is very similar to that observed in pure gauge non-compact QED when
{\em only} pure percolation occurs.
\figPeak

{\em This work has been partially supported by
research project CICYT AEN95/0882.}

\end{itemize}

\end{document}